# Band Gap of β-PtO$_2$ from First-principles

Yong Yang[1], Osamu Sugino[1,2], and Takahisa Ohno[1,3]

[1]*Global Research Center for Environment and Energy based on Nanomaterials Science (GREEN), National Institute for Materials Science (NIMS), Tsukuba, 305-0047, Japan*

[2]*Institute for Solid State Physics, University of Tokyo, 5-1-5 Kashiwanoha, Kashiwa 277-8581, Japan*

[3]*Computational Materials Science Unit, National Institute for Materials Science (NIMS), Tsukuba, 305-0047, Japan*

We studied the band gap of β-PtO$_2$ using first-principles calculations based on density functional theory (DFT). The results are obtained within the framework of generalized gradient approximation (GGA), GGA+U, *GW* and the hybrid functional method. For different types of calculations, the calculated band gap increases from ~ 0.46 eV to 1.80 eV. In particular, the band gap by *GW* (conventional and self-consistent) calculation shows a tendency of converging to ~ 1.25 ± 0.05 eV. Effect of the on-site Coulomb interaction on the bonding characteristics is also analyzed.

In recent years platinum dioxides have attracted considerable attention as a catalytic material for chemical reactions on surface/interface [1-8]. Actually, $PtO_2$ is known as Adam`s Catalyst [9, 10] in organic synthesis for a long time. For the oxidation of molecules on oxide surface, experiments have shown that the existence of $PtO_2$ can enhance the catalytic oxidation of carbon monoxide (CO) molecules [1], whose mechanism was theoretically explained later [2, 3]. In contrast to CO, the oxidation of NO molecules on $PtO_2$ (110) was theoretically predicted to be in a lower activity with comparison to that on pure Pt(111) surface [5]. The thin film of platinum oxide is observed to widely present on platinum electrode during electrochemical process [11], especially in the application of fuel cell [7, 8], which is an environmental friendly power resource. However, the effect of $PtO_2$ on the electrochemical processes, such as the oxygen reduction reaction (ORR) on Pt electrode, is poorly understood. As a first step, one should learn about the electronic properties of the material itself, which are, however, not yet well studied.

Experimentally, there are two crystalline phases of $PtO_2$ [12]: α-$PtO_2$ and β-$PtO_2$. The crystal of β-$PtO_2$ adopts the orthorhombic $CaCl_2$-type structure, which is distinguished from the other group VIII metal dioxides that have a tetragonal rutile structure [13]. Using density functional theory (DFT) calculations within the local density approximation (LDA), Wu and Weber explained the stability of the β-$PtO_2$ structure as originating from the strong hybridization between the Pt 5$d$ and O 2$p$ states around the Fermi level [14]. However, their calculated electronic density of states (DOS) indicates a metallic character of β-$PtO_2$, which contradicts with the experimental measurement that β-$PtO_2$ is a semiconductor with a resistivity of ~ $10^6$ Ω·cm [15]. Until now, no experimental data is available for the band gap of β-PtO$_2$, while the reported band gap of α-PtO$_2$ varies from ~ 1.3 eV [16] to 1.8 eV [17]. Our recent DFT-GGA calculation [18] with the PBE-type functional [19] gives a value of ~ 0.46 eV for β-PtO$_2$ band gap, which is further increased to be ~ 1.2 eV by using GGA+U scheme. The value of parameter U employed in the GGA+U calculation is justified [18] by using the conventional *GW* ($G_0W_0$) method [20, 21].

In this work, we go on to present a more comprehensive study on the band gap of β-PtO$_2$. In addition to GGA, GGA+U, and $G_0W_0$, we have performed the self-consistent *GW* and the Heyd-Scuseria-Ernzerhof (HSE) type hybrid functional [22-24] calculations to make a comparison of the results of different methods. The calculations are carried out by the Vienna *ab initio* simulation package (VASP) [25, 26], using a plane wave basis set and the projector augmented-wave (PAW) potentials [27, 28]. The exchange-correlation interactions of valence electrons are described by the PBE-type functional [19] for the GGA calculations, and by the HSE03 [22, 23] and HSE06 [24] functionals for the hybrid functional calculations. While doing the GGA+U calculations, the on-site Coulomb repulsion is treated using the simplified scheme introduced by Dudarev *et al.* [29]. As for the *GW* calculations, we take both the conventional and the self-consistent *GW* approach implemented in VASP [30, 31] to calculate the energy spectrum of quasiparticle. The quasiparticle energies and wave functions are obtained by solving the following equation [21]:

$$(T + V_{ext} + V_H)\psi_{nk}(\vec{r}) + \int d\vec{r}\,' \Sigma(\vec{r},\vec{r}\,';E_{nk})\psi_{nk}(\vec{r}\,') = E_{nk}\psi_{nk}(\vec{r}),$$

where $T$ is the kinetic energy operator of electrons, $V_{ext}$ is the external potential due to ions, $V_H$ is the electrostatic Hartree potential, $\Sigma$ is the electron self-energy operator, and $E_{nk}$ and $\psi_{nk}(\vec{r})$ are the quasiparticle energies and wave functions, respectively. Within the *GW* approximation, the self-energy operator $\Sigma$ is expressed as follows:

$$\Sigma(\vec{r},\vec{r}';E) = \frac{i}{2\pi}\int d\omega e^{i\delta\omega} G(\vec{r},\vec{r}';E+\omega) W(\vec{r},\vec{r}';\omega),$$

where $G$ is the Green's function, $W$ is the screened Coulomb interaction, and $\delta$ is a positive infinitesimal. For the convenience of discussion, we take the commonly used symbol "$G_0W_0$" for denoting the conventional *GW*, and "$GW_0$" for the partially self-consistent *GW*, where the self-consistent iteration is done for *G* only, and "*GW*" for the fully self-consistent *GW* approach, where the self-consistent iteration is done for both *G* and *W*.

All the calculations are done for the primitive cell of β-PtO$_2$, which contains 2 Pt atoms and 4 O atoms. In both GGA and GGA+U calculations, we use a 16×16×16 k-mesh and an energy cutoff of 600 eV for plane waves. In the HSE-type hybrid functional calculations, a 8×8×8 k-mesh and an energy cutoff of 600 eV for plane waves are employed. In the *GW* calculations, because of the heavy computational burden, the employed k-meshes range from 2×2×2 to 6×6×6, and the energy cutoff for plane waves range from 300 eV to 800 eV, to test the convergence of obtained band gap. The k-meshes are generated by the Monkhorst-Pack scheme [32]. The tetrahedron method with Blöchl corrections [33] is used for doing integral in the Brillouin zone (BZ).

The primitive cell of β-PtO$_2$ is shown Fig. 1(a), in which every Pt atom is coordinated by six O atoms, and accordingly every O atom is coordinated by three Pt atoms. The calculated band gap, and the optimized length of the orthorhombic cell edges *a*, *b*, and *c* with the GGA+U method are shown in Fig. 1(b) and Fig. 1(c), respectively, as a function of the effective on-site Coulomb repulsion ($U_{eff}$). The data points at which $U_{eff}$ = 0 correspond to the pure DFT-GGA calculations. The band gap shows a general tendency of increasing with enlarged $U_{eff}$, and the calculated cell parameters *a* and *c* show a trend of decreasing with enlarged $U_{eff}$ while the length of cell edge *b* fluctuates with $U_{eff}$. At the point where $U_{eff}$ = 7 eV, there is a slight dip in the value of band gap. This is related to the structural relaxation of atoms and cell geometries. As shown in Fig. 1(c), the length of cell edge *c* increases slightly at $U_{eff}$ = 7 eV with comparison to the value of *c* at $U_{eff}$ = 6.5 eV, which is out of the general variation trend of dropping. When the atomic positions and cell parameters are kept the same for the calculations with different $U_{eff}$, we find that the dipping point vanishes and the band gap increases monotonically with increasing $U_{eff}$. The relaxed cell parameters by GGA+U calculations with the value of $U_{eff}$ of about 7.5 eV match the experimental values best. In the following paragraphs, we will show that the calculated band gap with $U_{eff}$ = 7.5 eV also agrees quite well with *GW* calculations.

Figure 2(a) shows the calculated electronic density of states (DOS) of β-PtO$_2$ using the conventional and self-consistent *GW* method ($G_0W_0$, $GW_0$, *GW*). Compared to the GGA band gap (0.46 eV) and the GGA+U band gap (1.20 eV) in our recent work [18], the calculated band gap by $G_0W_0$, $GW_0$, and *GW* is 1.31 eV, 1.45 eV, and 1.50 eV, respectively. A 4×4×4 k-mesh and an energy cut-off of 600 eV for plane waves and totally 400 energy bands are used for the $G_0W_0$, $GW_0$, and *GW* calculations to obtain the data in Fig. 2(a). Calculated GGA energy bands along some symmetry lines in the BZ are shown in Fig. 2(b), together with some $G_0W_0$ quasiparticle energy points. It is found that the energy dispersion of GGA and the $G_0W_0$ energy points is similar. Due to the usage of a smaller k-mesh, the GGA band gap deduced from Fig. 2(b) is ~ 0.7 eV, which indicates that the k-points corresponding to the band gap of ~ 0.46 eV in the more accurate DOS calculation are missing from the high symmetry lines. The key difference between GGA and $G_0W_0$ is that the $G_0W_0$ energies of the conduction bands are sitting at higher values. The partial DOS (PDOS) of Pt 5$d$ and O 2$p$ orbitals from GGA and GGA+U calculations are displayed in Figs. 2(c) and 2(d), respectively. Around the Fermi level, strong hybridization is found between the Pt 5$d$ and O 2$p$ orbitals in the GGA calculations. The Pt 5$d$ and O 2$p$ components in the bonding states near the Fermi level are almost the same for GGA. This implies that electron transfer can take place between the Pt 5$d$ and O 2$p$ orbitals with equal probability and magnitude. It is possible that the strong $pd$ hybridization originates from the general trend that LDA/GGA insufficiently describes the interactions between the $d$ electrons. Indeed, the GGA+U calculation provides qualitatively different DOS where the center of the Pt 5$d$ bands is shifted downward by ~ 4 eV. This is presumably due to the Coulomb interaction of $d$ electrons. As a result, the $pd$ hybridization that appears in GGA is largely reduced, and the O 2$p$ orbitals contribute to the major part of the electron DOS at energies ~ 2 to 3 eV below the Fermi level (Fig. 2(d)). Note that in ZnS [34] and ZnO [35] the $d$-bands were reported to be similarly shifted downward when the Coulomb interaction was incorporated by using the GGA+U method.

Figure 3 shows the DOS of β-PtO$_2$ calculated by using the HSE03 and HSE06 hybrid functionals. Below the Fermi level, the DOS from HSE03 and HSE06-type calculations are almost the same. The features of the DOS above the Fermi level are also similar except that the bottom of conduction bands differs by ~ 0.2 eV. The calculated band gap is ~ 1.60 eV by HSE03 and is ~1.81 eV by HSE06. This indicates that, the HSE band gap depends notably on the empirical parameter $\mu$ [36] that defines the short-range and long-range Coulomb interactions in the exchange-correlation functional. The characteristic distance separating the short and long-range Coulomb interactions is $L_c = 2/\mu$, where $\mu = 0.3$ Å$^{-1}$ for HSE03 and $\mu = 0.2$ Å$^{-1}$ for HSE06. It follows that, $L_c \approx 6.67$ Å for HSE03 and $L_c = 10$ Å for HSE06.

To test the convergence of the computed $G_0W_0$ band gap, we have done calculations with varying energy cut-off of plane waves and k-meshes. Shown in Fig. 4(a), is the calculated electronic DOS diagrams at different energy cut-off using a 4×4×4 k-mesh and 400 energy bands for the summation of quasiparticle energies. When the energy cut-off increases from 300 eV to 800 eV, the band gap is kept at ~ 1.3 eV, and the DOS features near the Fermi level is only slightly modified. That means, the energy cut-off of ~ 300 eV is enough for the $G_0W_0$ calculation of band gap. The variation of DOS with different k-meshes is shown in Fig. 4(b). The band gap varies from ~ 1.30 eV (4×4×4 k-mesh), and ~ 1.25 eV (5×5×5 k-mesh, run1 & run2), to ~ 1.22 eV (6×6×6 k-mesh). Calculation with finer k-meshes is not able to carry out, due to the quota of available computational resource. In spite of this limitation, the calculated $G_0W_0$ band gap shows a tendency of converging to ~ 1.2 eV. The partially self-consistent $GW_0$ gives a band gap of ~ 1.42 eV and 1.28 eV, for calculations using a 4×4×6 and 6×6×6 k-mesh, respectively. For a wide range of materials, the $GW_0$ usually give the best value of band gap comparing to experiment [31]. The self-consistent *GW* band gap is ~ 1.46 eV by calculation using a 4×4×6 k-mesh. Since the $GW_0$ and *GW* band gap is close to each other in calculations with the same k-mesh, we expect that the self-consistent *GW* band gap will converge to ~ 1.30 eV when calculated using a 6×6×6 or denser k-meshes. In the following paragraphs, we will go on study the convergence of the $G_0W_0$, which is computationally less demanding than the partially self-consistent $GW_0$ and the self-consistent *GW* method.

The convergent behavior of $G_0W_0$ calculation with respect to the number of bands can be very different from one system to another [30]. Recent studies on ZnO [35, 37] have shown that, the $G_0W_0$ band gap does not converge until a vastly large number of energy bands are included in the calculation. We have also made calculations to verify this point. Figure 5 shows the $G_0W_0$ band gap as a function of k-meshes and the number of energy bands employed in the calculation. For the 2×2×2 and 3×3×3 k-meshes, the calculated band gap fluctuates around 1.61 eV and 1.47 eV when the number of bands (NBAND) increases from 200 to 800. The band gap of 4×4×4 k-mesh decreases continuously with increasing number of bands, from ~ 1.33 eV (NBAND = 200) to 1.27 eV (NBAND = 800). In the case of both 5×5×5 and 6×6×6 k-mesh, the parameter NBAND varies from 216 to 720. The calculated band gap fluctuates at ~ 1.25 eV (~ 1.26 to ~ 1.23 eV) for the 5×5×5 k-mesh and at ~ 1.22 eV (~ 1.24 to ~ 1.22 eV) for the 6×6×6 k-mesh. From the data lines shown in Fig. 5, we expect that the band gap will converge to ~ 1.2 eV when a much larger number of bands and k-points are employed in the $G_0W_0$ calculation.

Within the $G_0W_0$ approach, usually the quasiparticle wave function is very close to the LDA/GGA wave function, therefore the quasiparticle energy ($E_{nk}^{QP}$) can be calculated by using only the diagonal matrix element of the self-energy operator [21]. As a result, the quasi-particle equation is reduced to [30]

$$E_{nk}^{QP} = Re[\langle\psi_{nk}|T + V_{ext} + V_H + \Sigma(E_{nk}^{QP})|\psi_{nk}\rangle],$$

where the term $\Sigma(E_{nk}^{QP})$ is calculated using the LDA/GGA Kohn-Sham eigenvalues and eigenfunctions, and the eigenstates $\psi_{nk}$ are approximated by the LDA/GGA wave functions. Thus, the convergence of the DFT-GGA wave functions will guarantee the convergence of the quasi-particle energy and consequently the value of band gap. Figure 6 shows the DFT-GGA total energy of β-PtO$_2$ as a function of k-mesh, for the calculations with a plane waves' energy cut-off of 400 eV and 600 eV, respectively. In both cases, the total energy is well converged (≤ 2 meV) when the employed k-mesh is equal to or denser than 6×6×6. Within the framework of density functional theory (DFT), the total energy of a system is solely determined by the electron density. The convergence of total energy indicates the convergence of electron density, or alternatively the wave function of the system (differs by a phase factor at most). This result in return suggests that the calculated $G_0W_0$ band gap by using a 6×6×6 k-mesh is close to convergence.

From the experimental side, the measured band gap of amorphous PtO$_2$ is ~ 1.2 eV [38], and that of α-PtO$_2$ ranges from ~ 1.3 eV [16] to 1.8 eV [17]. No experimental data is available for the band gap of β-PtO$_2$ yet. The reported resistivity for α-PtO$_2$ is on the order of 10$^4$ Ω·cm [39, 40] and that for β-PtO$_2$ is ~ 10$^6$ Ω·cm [15]. Although the value of band gap is not the only factor that determines the resistivity of a material, it would be acceptable that the alpha and beta form of PtO$_2$ have similar band gap when considering their electrical resistivity.

In conclusion, we have studied the band gap of β-PtO$_2$ by DFT calculations. DFT-GGA calculation gives a band gap of ~ 0.46 eV, which is usually underestimated. The band gap is increased when the effective on-site Coulomb repulsion (U$_{eff}$) between the Pt 5*d* electrons is taken into account via the GGA+U method. In the case U$_{eff}$ ~ 7.5 eV, which yields the best fit to the experimental lattice parameters, the calculated band gap is ~ 1.2 eV. The band gap given by conventional $G_0W_0$ approach is ~ 1.25 ± 0.05 eV, which shows a tendency of converging to ~ 1.2 eV when the number of energy bands, the k-meshes and the energy cut-off for plane waves employed in calculation are increased to large values. The partially self-consistent $GW_0$ gives a band gap of ~ 1.28 eV and the self-consistent *GW* band gap is ~ 1.46 eV, which is expected to converge to a lower value (likely ~ 1.30 eV) when more accurate calculations are performed. The band gap given by HSE03 and HSE06 hybrid functional calculation is ~ 1.60 eV and 1.81 eV, respectively. The *GW* and HSE calculations indicate that the value of U parameter used in our recent work on oxygen-deficient β-PtO$_2$ [18] belongs to the range U ≥ 7.5 eV. We hope that the predicted band gap can be tested by optical measurement in the future.

**Acknowledgments**

This work is supported by the Global Research Center for Environment and Energy based on Nanomaterials Science (GREEN) at National Institute for Materials Science (NIMS). The first-principles calculations were carried out by the supercomputer (SGI Altix) of NIMS.

**Figures and Captions**

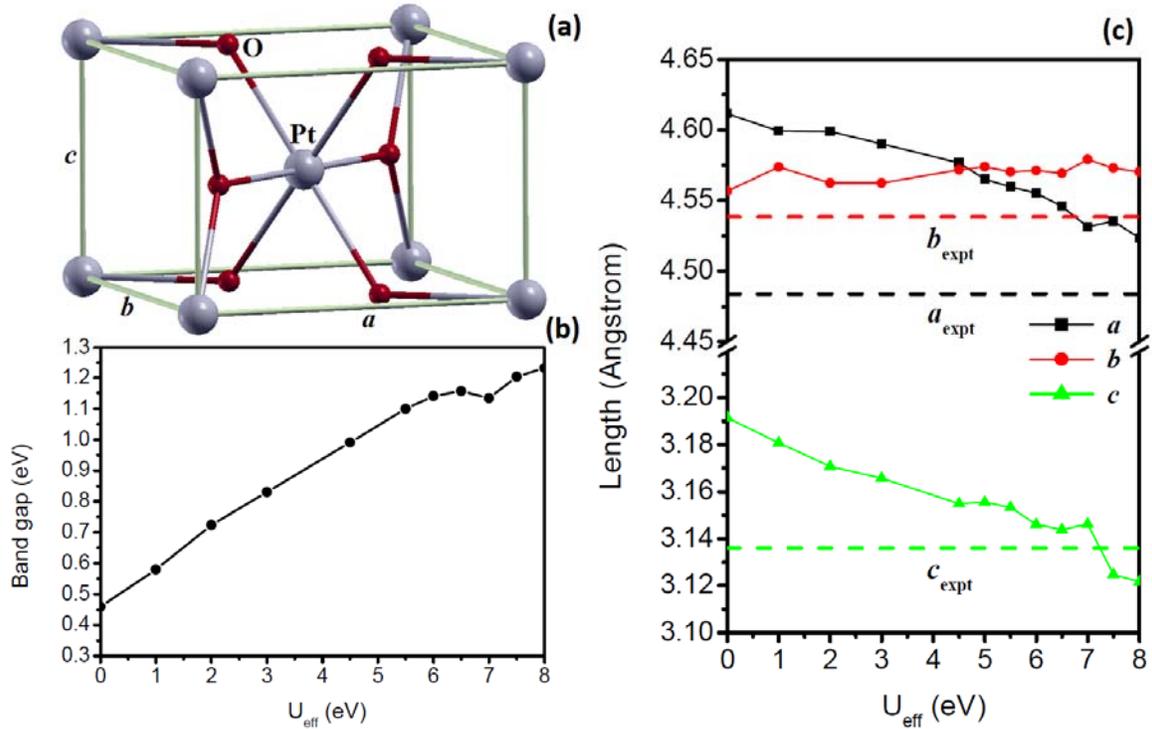

**FIG. 1 (Color online)** (a) Primitive cell of β-PtO$_2$. (b) Calculated band gap of β-PtO$_2$ as a function of effective on-site Coulomb repulsion (U$_{eff}$) in GGA+U method. (c) Optimized length of cell edges *a*, *b*, and *c* as a function of U$_{eff}$. The experimental values of *a*, *b*, and *c* (from Ref. [13]) are indicated by dashed lines.

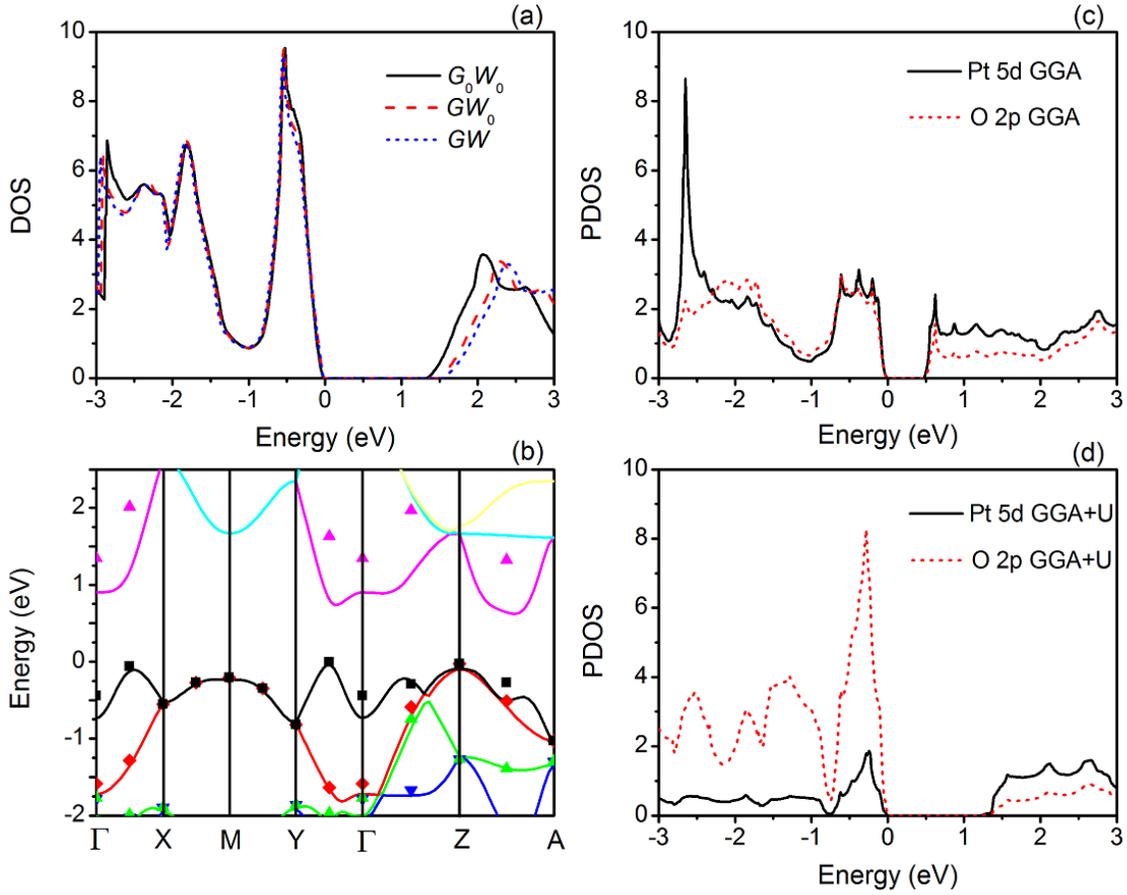

**FIG. 2 (Color online)** (a) Calculated electronic density of states (DOS) of β-PtO$_2$ by using one-shot $G_0W_0$, partially self-consistent $GW_0$ and self-consistent $GW$ method. (b) Energy bands of β-PtO$_2$, calculated using GGA (solid lines) and $G_0W_0$ method (scattered squares and triangles) along some lines joining the high-symmetry points in the k-space. The direct coordinates of the k-points in Brillouin zone (BZ): Γ = (0, 0, 0), X = (0.5, 0, 0), M = (0.5, 0.5, 0), Y = (0, 0.5, 0), Z = (0, 0, 0.5), A = (0.5, 0.5, 0.5). (c) The partial DOS (PDOS) of Pt 5*d* and O 2*p* orbitals from GGA calculations. (d) The PDOS of Pt 5*d* and O 2*p* orbitals from GGA+U calculations. For all the figures here and below, the highest occupied energy level is set at zero and the unit of DOS and PDOS is state/eV/cell.

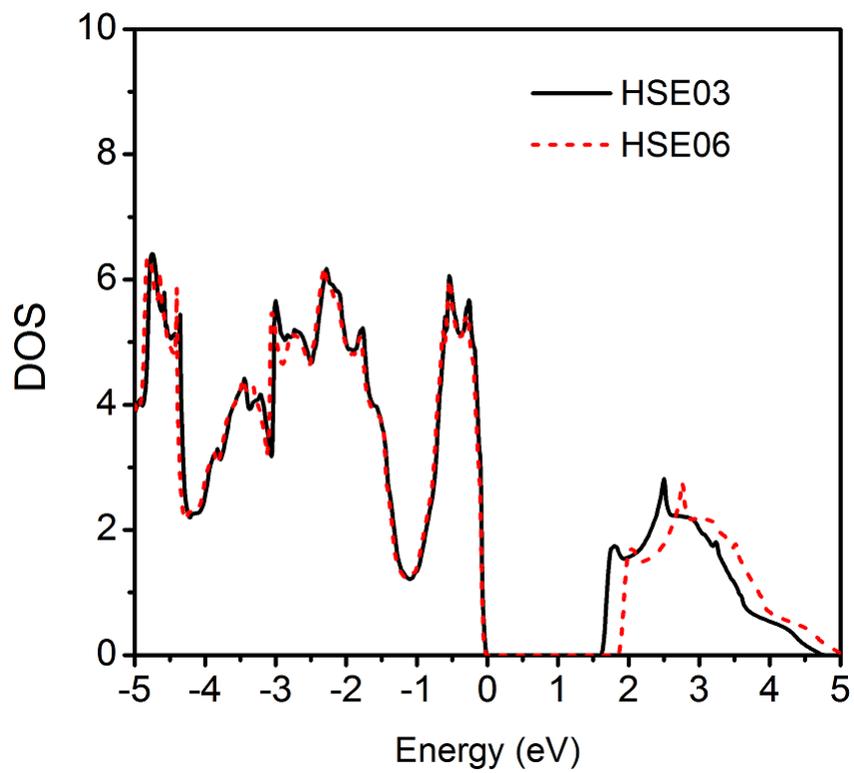

FIG. 3 (Color online) The electronic DOS of β-PtO$_2$, obtained from HSE03 and HSE06 hybrid functional calculations.

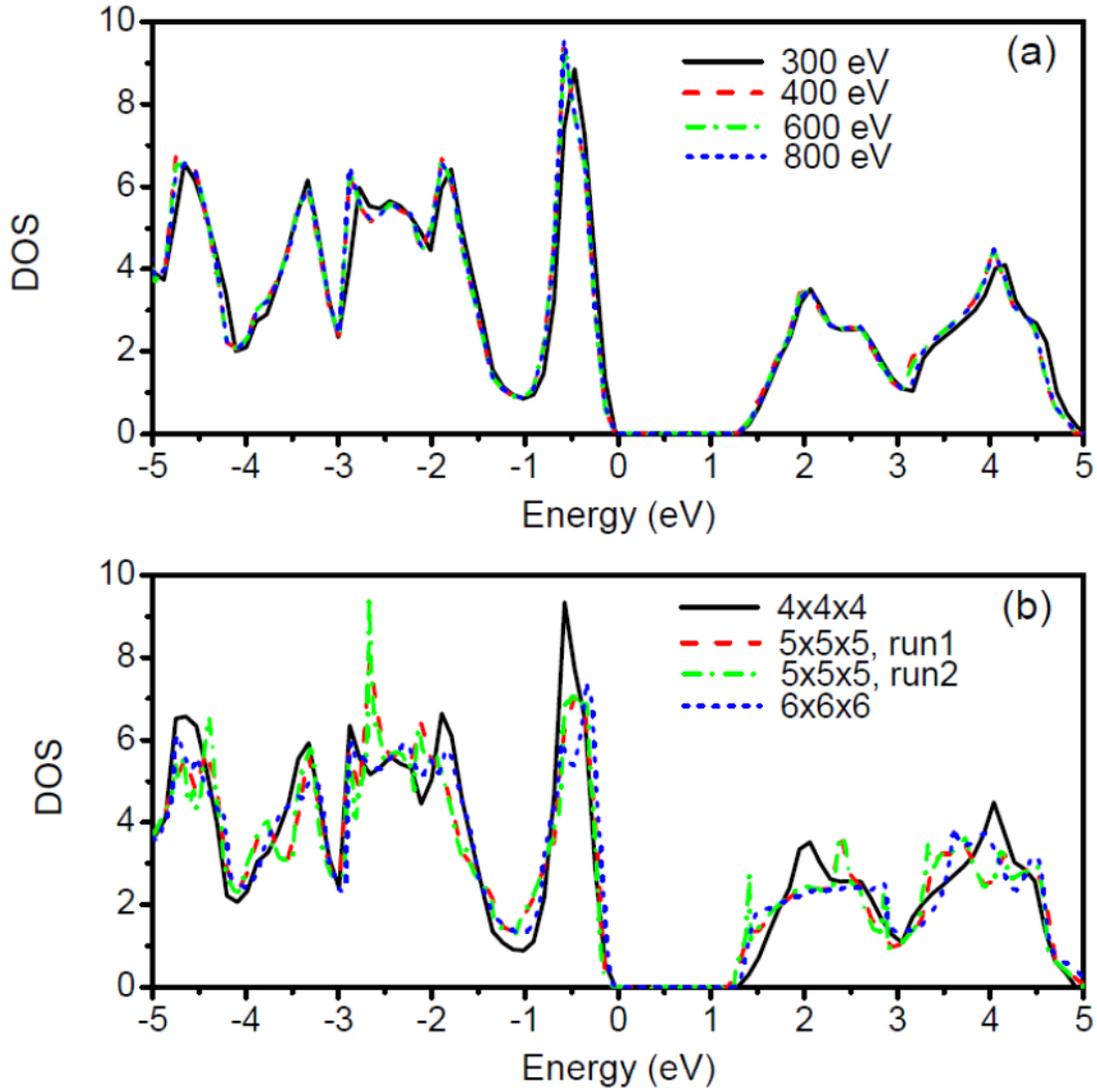

**FIG. 4 (Color online)** The $G_0W_0$ electronic DOS of β-$PtO_2$. (a) Results obtained from four calculations with different energy cut-off for plane waves. A 4×4×4 k-mesh and 400 energy bands are used for each calculation. (b) Results from the calculations using four different k-meshes. The energy cut-off (ENCUT) and number of energy bands (NBAND) for both k-meshes 4×4×4 and 5×5×5 run1 are 600 eV and 400. The ENCUT and NBAND for both k-meshes 5×5×5 run2 and 6×6×6 are 400 eV and 450.

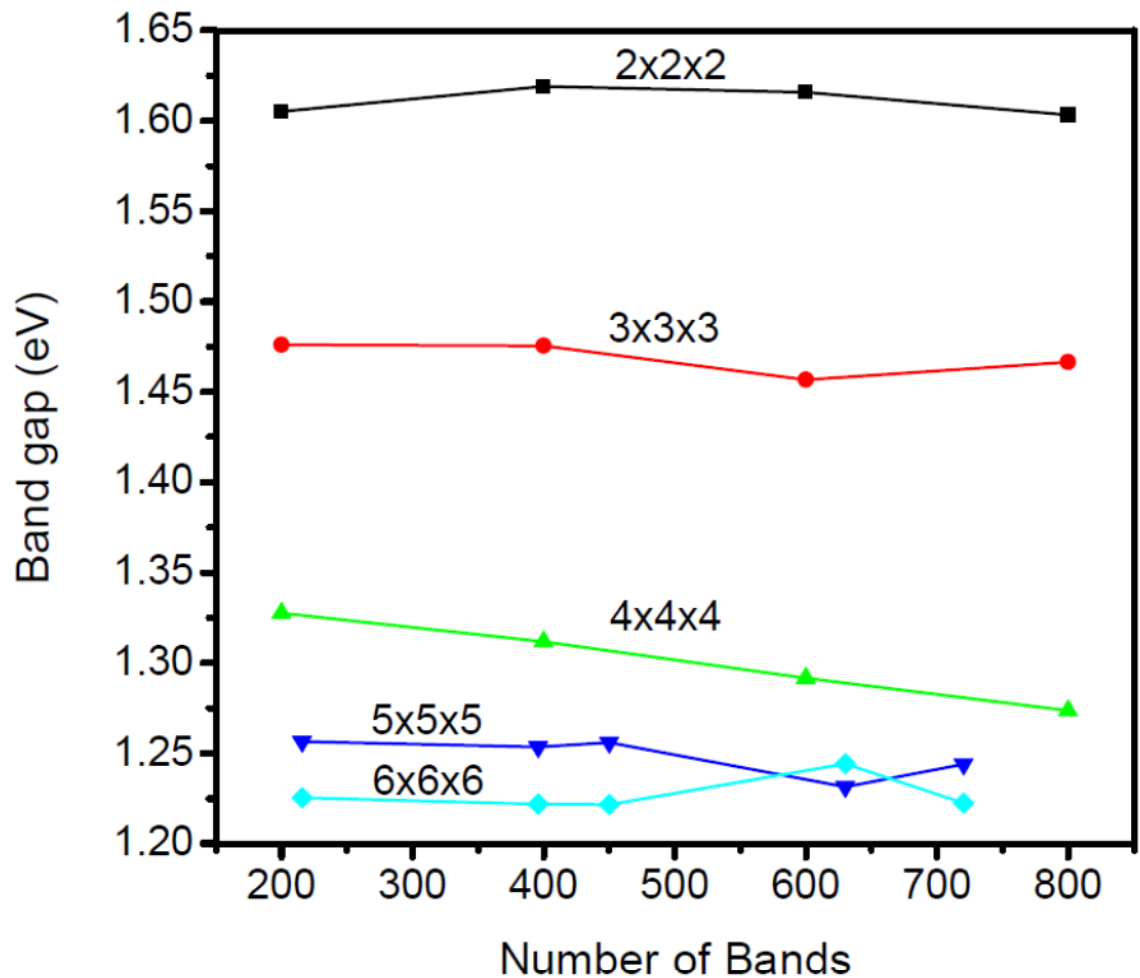

**FIG. 5 (Color online)** The $G_0W_0$ band gap of β-$PtO_2$, as a function of the number of energy bands and k-meshes involved in calculation. The plane waves' energy cut-off is 600 eV for k-meshes 2×2×2, 3×3×3, 4×4×4 and 5×5×5, and is 400 eV for the k-mesh 6×6×6.

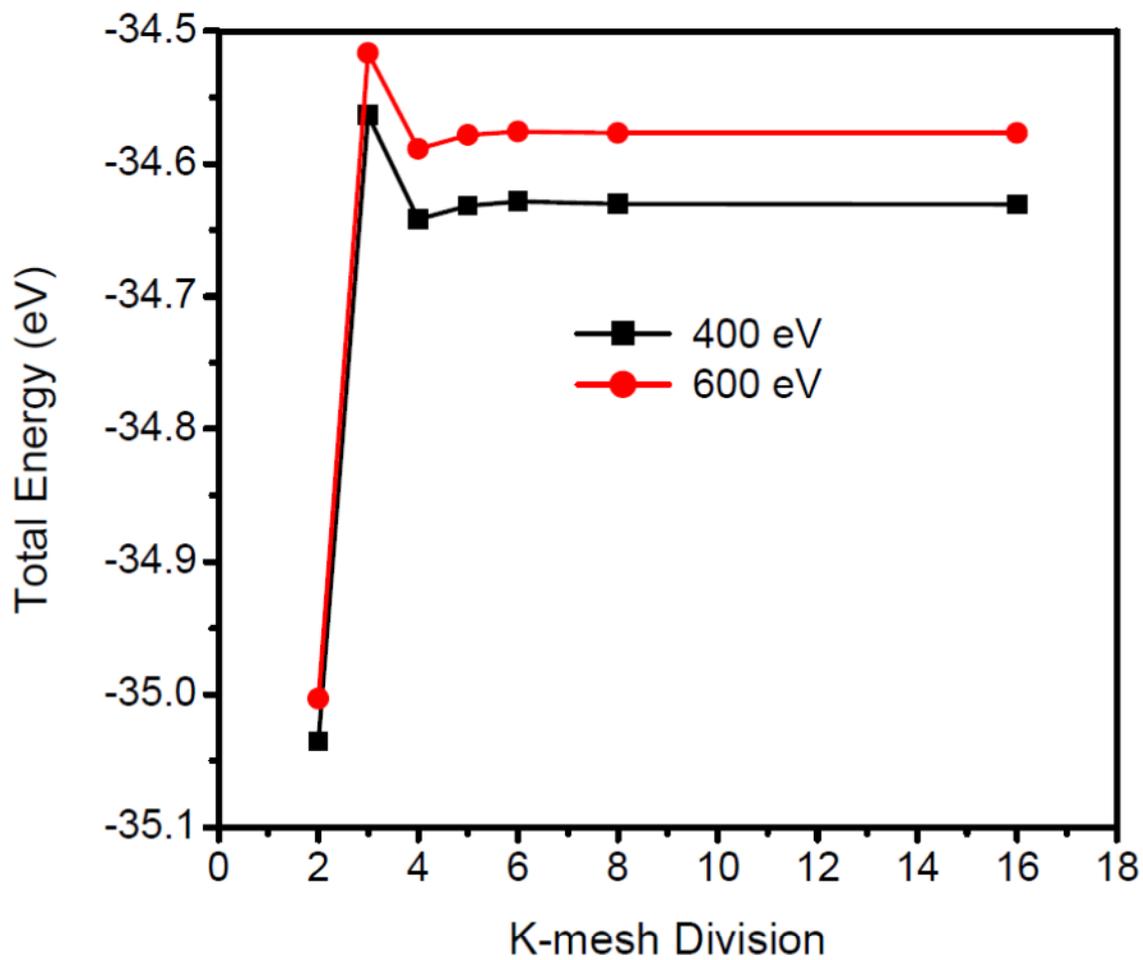

**FIG. 6 (Color online)** The DFT-GGA total energy of β-PtO$_2$, as a function of k-meshes employed in two calculations with a plane waves' energy cut-off of 400 eV and 600 eV. Each integer *n* on the k-mesh division axis corresponds to a uniform k-mesh of *n×n×n*.